\documentclass{article}
\usepackage{spconf}

\title{Spectral or Spatial? Leveraging Both for Speaker Extraction in Challenging Data Conditions}

\name{Aviad Eisenberg$^{1,2}$, Sharon Gannot$^{1}$, Shlomo E. Chazan$^{2}$
\thanks{This project has received funding from the ‘Audition’ Project, Data Science Program, Council of Higher Education, Israel.}}
\address{
$^{1}$Faculty of Engineering, Bar-Ilan University, Ramat-Gan, Israel\; $^{2}$OriginAI, Israel
}

\usepackage{cite}
\usepackage{graphicx}
\usepackage{comment}
\usepackage{balance}
\usepackage{acronym}
\usepackage{amsmath,amssymb}
\usepackage{tabularx}
\usepackage[hang]{subfigure}
\usepackage{booktabs}
\usepackage{tikz}
\usetikzlibrary{positioning}
\usepackage{caption}
\usetikzlibrary{shapes.multipart}
\usepackage{amsfonts}
\usepackage{enumitem,bbm}
\usepackage{wrapfig}
\usepackage[normalem]{ulem}
\usepackage[utf8]{inputenc}
\usepackage[normalem]{ulem}

\usetikzlibrary{shapes.geometric, arrows}

\def\x{\mathbf{x}(t,k)}
\def\sq{s_q(t,k)}
\def\sd{s_d(t,k)}
\def\si{s_i(t,k)}
\def\hq{\mathbf{h}_q(t,k)}

\def\hn{\mathbf{h}_n(t,k)}
\def\td{\theta_d}

\acrodef{STFT}{short-time Fourier transform}
\acrodef{ISTFT}{inverse short-time Fourier transform}
\acrodef{BSS}{blind source separation}
\acrodef{DOA}{direction of arrival}
\acrodef{DC}{deep clustering}
\acrodef{DPRNN}{dual-path recurrent neural network}
\acrodef{TF}{time-frequency}
\acrodef{TCN}{temporal convolutional network}
\acrodef{ATF}{acoustic transfer function}
\acrodef{SI-SDR}{scale-invariant signal-to-distortion ratio}
\acrodef{MSE}{mean square error}
\acrodef{DFT}{discrete Fourier transform }
\acrodef{SIR}{signal-to-interference ratio}
\acrodef{OVA}{overlap-and-add}
\acrodef{SDR}{signal-to-distortion ratio}
\acrodef{BLSTM}{Bidirectional Long Short-Term Memory}
\acrodef{SOTA}{state-of-the-art}
\acrodef{RI}{Real-Imaginary}
\acrodef{RIR}{room impulse response}
\acrodef{SNR}{signal-to-noise ratio}
\acrodef{RNN}{Recurrent Neural Networks}
\acrodef{FLOP}{floating point operation}
\acrodef{WPE}{weighted prediction error}
\acrodef{CASA}{computational simultaneous grouping scene analysis}
\acrodef{FC}{fully connected}
\acrodef{E2E}{end-to-end}
\acrodef{SDR}{signal to distortion ration}
\acrodef{SIR}{signal-to-interference ratio}
\acrodef{STOI}{short-time objective intelligibility}
\acrodef{PESQ}{perceptual evaluation of speech quality}
\acrodef{CNN}{convolutional neural network}
\acrodef{MVDR}{minimum variance distortion beamformer}
\acrodef{RTF}{relative transfer function}
\acrodef{IPD}{inter-channel phase difference}
\acrodef{LPS}{logarithm power spectra}
\acrodef{AF}{angle features}
\acrodef{DSB}{delay-and-sum beamforme}
\acrodef{SI-SDR}{scale-invariant signal-to-distortion ratio}
\acrodef{FiLM}{feature-wise linear modulation}
\acrodef{ULA}{uniform linear array}
\acrodef{TPD}{target phase difference}

\tikzset{
  block/.style={
    rectangle,
    rounded corners,
    draw=black,
    very thick,
    minimum height=3em,
    minimum width=3em
  }
}
\tikzset{
  input/.style={
    circle,
    draw=black,
    very thick,
    minimum size=2em
  }
}
\tikzset{
  output/.style={
    circle,
    draw=black,
    very thick,
    minimum size=2em
  }
}

\begin{document}
\ninept
\maketitle

\begin{abstract}
This paper presents a robust multi-channel speaker extraction algorithm designed to handle inaccuracies in reference information. While existing approaches often rely solely on either spatial or spectral cues to identify the target speaker, our method integrates both sources of information to enhance robustness. A key aspect of our approach is its emphasis on stability, ensuring reliable performance even when one of the features is degraded or misleading. Given a noisy mixture and two potentially unreliable cues, a dedicated network is trained to dynamically balance their contributions—or disregard the less informative one when necessary. We evaluate the system under challenging conditions by simulating inference-time errors using a simple \ac{DOA} estimator and a noisy spectral enrollment process. Experimental results demonstrate that the proposed model successfully extracts the desired speaker even in the presence of substantial reference inaccuracies.
\end{abstract}

\section{Introduction}
The task of extracting a target speaker from a mixture of speakers has been extensively studied in the literature, but remains a significant challenge \cite{zmolikova2023neural}. Unlike the general speech separation task, this problem focuses on isolating a specific speaker from a speech signal that is overlapping with others. The critical factor in extracting the desired speaker lies in the availability of relevant enrollment information.

In the single-channel case, such as in \cite{eisenberg2022single,eisenberg2023two,ge2020spex+,wang2018voicefilter}, the model performs the extraction task based solely on the spectral information in the mixture. 
In scenarios involving microphone arrays, spatial information becomes available. Algorithms designed for such cases can be categorized into spectral-based and spatial-based. Spectral-based approaches rely solely on spectral features of the enrollment signal. For example, in \cite{ge2022spex}, a single-channel reference signal is used for both extracting the target speaker and estimating its \ac{DOA}. This method employs a masking operation to construct a \ac{MVDR} beamformer directed toward the target speaker. The use of spatial features derived from the mixture signal is explored in \cite{zorilua2021investigation}. Additionally, the well-known SpeakerBeam model \cite{vzmolikova2019speakerbeam} is extended in \cite{delcroix2020improving} by incorporating spatial features from the mixture signal, demonstrating improved performance. Although the aforementioned studies consider spatial information within the mixture, the enrollment signal guiding the model toward the target speaker remains a single-channel reference signal devoid of spatial characteristics.

In contrast to spectral-based speaker extraction algorithms, where the identification of the desired speaker relies solely on spectral information extracted from the enrolment signal, other approaches incorporate spatial cues, such as the desired \ac{DOA}, to enhance performance. For instance, \cite{tesch2023spatially} highlights the advantages of utilizing directional reference features. Similarly, \cite{gu2020temporal} proposes a method that combines spectral features, such as \ac{LPS}, with spatial features like \ac{IPD} and \ac{AF}, to generate a mask guided by the \ac{DOA} of the target speaker. 
In \cite{elminshawi2023beamformer}, a \ac{DSB} is applied using the \ac{DOA} to enhance the target speaker's signal. The single-channel output of the beamformer is then fed into an auxiliary network, which generates a time-varying vector used as a reference for the main extraction network. 
In \cite{gu2024rezero}, the model was trained to identify speakers located within predefined spatial regions. Three types of spatial constraints were considered: \ac{DOA}-based regions defining angular sectors, radius-based regions forming circular areas, and combined \ac{DOA} and radius constraints delineating conical volumes. This approach represents a generalization beyond traditional spatial-based speaker extraction methods since it extracts all sources existing in the given region. In this paper, the direction feature is derived from the similarity between \ac{IPD} and \ac{TPD} within the query angle window. At the same time, a distance embedding generator is learned for the radius query.


Several studies have investigated the integration of auxiliary utterances to improve speaker extraction. For example, in \cite{martin2019multi}, an additional audio sample from the same location as the desired speaker is assumed to be available. Using this sample, the SpeakerBeam framework estimates a mask that is then used to compute an \ac{MVDR} beamformer. However, the specific contributions of the auxiliary audio and spatial features to the separation process remain unclear. Moreover, when the speakers in the mixture are spatially close to each other, the performance is expected to degrade significantly due to the limitations of the beamforming approach.
Other studies have also incorporated visual features, such as videos, as auxiliary information \cite{gu2020multi, xu2020neural, li2023audio, xu2023multi}. For instance, \cite{gu2020multi} presents a framework that combines spectral, spatial, and video-based references to enhance speaker extraction. This study also highlights the performance degradation when only spectral-spatial features are used, particularly in scenarios where the \ac{DOA} estimates include errors and the speakers in the mixture are relatively close to each other.
In \cite{eisenberg2025e2e}, a framework was introduced to evaluate various features for multichannel speaker extraction. Among the examined features were the \ac{DOA}, a single-channel reference signal, and the instantaneous \ac{RTF}.

To the best of our knowledge, none of the aforementioned approaches achieves sufficiently robust and stable performance in the presence of errors in the provided reference information, especially when reference information is partially missing. 
The contribution of our work is threefold:  
1) We propose a fully integrated approach that combines spectral and spatial features, demonstrating that both have an equal impact on the system's performance. This is achieved by incorporating a classification module during training and employing a dedicated training procedure designed to account for inaccurate data. 
2) We introduce a simple but practical procedure for \ac{DOA} estimation and alignment, ensuring each spectral reference is accurately matched to its corresponding \ac{DOA}.  
3) We address scenarios where one of the auxiliary inputs is faulty, showcasing the robustness of our model under such conditions.

\section{Problem Formulation}
A scenario involving $Q$ concurrent speakers, recorded by $J$ microphones in a reverberant and noisy environment, is considered. The problem is analyzed in the \ac{STFT} domain, where \( k \in \{0, \ldots, K-1\} \) and \( t \in \{0, \ldots, T-1\} \) denote the frequency and time-frame indices, respectively. Here, \( T \) and \( K \) represent the total number of time-frames and frequency bands. Let \( \sq \) represent the clean, anechoic speech signal of the \( q \)-th speaker.
The signal captured by the microphone array can be formulated as \begin{equation}
\scalebox{0.95}{$
\x = \sum_{q=1}^{Q}  \hq \cdot \sq   + \hn \cdot  n(t,k) ,
$}
\end{equation}
where $\hq$ is a $J\times 1$ vector of the \acp{ATF} relating the $q$-th source and the microphones array, $\hn$ is the \acp{ATF} relating the noise and the microphones array, and $n(t,k)$ is the anechoic additive noise.

We focus on a scenario with two concurrent speakers (\( Q=2 \)), referred to as the desired speaker, \( \sd \), and the interference speaker, \( \si \). 
The reverberant desired signal captured by the first microphone is denoted  $\tilde{s}_d(t,k)$.
The goal is to extract the reverberant desired speaker's signal, \( \hat{\tilde{s}}_d(t,k) \), from the mixed signal \( \x \), using a reverberant single-channel enrollment signal $\tilde{{e}}_d(t,k)$  
the desired speaker's \ac{DOA}, denoted as \( \theta_d \).

\begin{figure*}
    \centering
    
\scalebox{0.7}{ 

\begin{tikzpicture}[x=0.75pt,y=0.75pt,yscale=-1,xscale=1]

\draw  [fill={rgb, 255:red, 229; green, 226; blue, 189 }  ,fill opacity=1 ] (243.2,245.9) .. controls (243.2,236.9) and (250.5,229.6) .. (259.5,229.6) -- (366.9,229.6) .. controls (375.9,229.6) and (383.2,236.9) .. (383.2,245.9) -- (383.2,294.8) .. controls (383.2,303.8) and (375.9,311.1) .. (366.9,311.1) -- (259.5,311.1) .. controls (250.5,311.1) and (243.2,303.8) .. (243.2,294.8) -- cycle ;
\draw    (366,268.43) -- (331.28,268.47) ;
\draw [shift={(329.28,268.47)}, rotate = 359.94] [color={rgb, 255:red, 0; green, 0; blue, 0 }  ][line width=0.75]    (10.93,-3.29) .. controls (6.95,-1.4) and (3.31,-0.3) .. (0,0) .. controls (3.31,0.3) and (6.95,1.4) .. (10.93,3.29)   ;
\draw   (329.26,257.39) -- (264.07,257.48) -- (264.05,247.81) -- (329.25,247.72) -- cycle ;
\draw   (329.28,267.05) -- (264.08,267.15) -- (264.07,257.48) -- (329.26,257.39) -- cycle ;
\draw   (329.29,276.72) -- (264.09,276.81) -- (264.08,267.14) -- (329.28,267.05) -- cycle ;
\draw  [fill={rgb, 255:red, 245; green, 166; blue, 35 }  ,fill opacity=1 ] (329.3,286.39) -- (264.11,286.48) -- (264.09,276.81) -- (329.29,276.72) -- cycle ;
\draw   (329.32,296.05) -- (264.12,296.14) -- (264.11,286.48) -- (329.3,286.39) -- cycle ;
\draw   (329.25,247.72) -- (264.05,247.81) -- (264.04,238.14) -- (329.24,238.05) -- cycle ;

\draw  [fill={rgb, 255:red, 237; green, 228; blue, 228 }  ,fill opacity=1 ] (373.5,154.5) .. controls (373.5,148.98) and (377.98,144.5) .. (383.5,144.5) -- (493.5,144.5) .. controls (499.02,144.5) and (503.5,148.98) .. (503.5,154.5) -- (503.5,184.5) .. controls (503.5,190.02) and (499.02,194.5) .. (493.5,194.5) -- (383.5,194.5) .. controls (377.98,194.5) and (373.5,190.02) .. (373.5,184.5) -- cycle ;
\draw  [fill={rgb, 255:red, 204; green, 224; blue, 184 }  ,fill opacity=1 ] (10,246.3) .. controls (10,237.3) and (17.3,230) .. (26.3,230) -- (151.51,230) .. controls (160.51,230) and (167.81,237.3) .. (167.81,246.3) -- (167.81,295.2) .. controls (167.81,304.2) and (160.51,311.5) .. (151.51,311.5) -- (26.3,311.5) .. controls (17.3,311.5) and (10,304.2) .. (10,295.2) -- cycle ;
\draw (49.71,268.47) node  {\includegraphics[width=44.27pt,height=43.96pt]{ 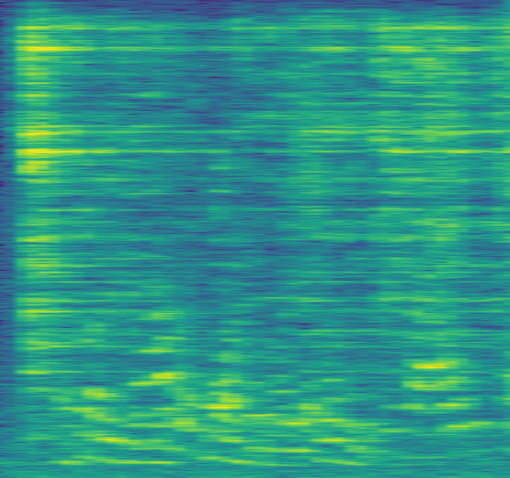}};
\draw   (20.19,239.16) -- (79.22,239.16) -- (79.22,297.78) -- (20.19,297.78) -- cycle ;

\draw   (90.83,233.89) -- (160.93,255.03) -- (160.89,285.57) -- (90.73,306.51) -- cycle ;
\draw    (242.84,271.05) -- (218.95,270.77) ;
\draw [shift={(216.95,270.75)}, rotate = 0.66] [color={rgb, 255:red, 0; green, 0; blue, 0 }  ][line width=0.75]    (10.93,-3.29) .. controls (6.95,-1.4) and (3.31,-0.3) .. (0,0) .. controls (3.31,0.3) and (6.95,1.4) .. (10.93,3.29)   ;
\draw    (167.67,270.94) -- (195,270.76) ;
\draw [shift={(197,270.75)}, rotate = 179.63] [color={rgb, 255:red, 0; green, 0; blue, 0 }  ][line width=0.75]    (10.93,-3.29) .. controls (6.95,-1.4) and (3.31,-0.3) .. (0,0) .. controls (3.31,0.3) and (6.95,1.4) .. (10.93,3.29)   ;
\draw   (197,270.75) .. controls (197,265.94) and (201.47,262.04) .. (206.98,262.04) .. controls (212.49,262.04) and (216.95,265.94) .. (216.95,270.75) .. controls (216.95,275.56) and (212.49,279.46) .. (206.98,279.46) .. controls (201.47,279.46) and (197,275.56) .. (197,270.75) -- cycle ; \draw   (197,270.75) -- (216.95,270.75) ; \draw   (206.98,262.04) -- (206.98,279.46) ;
\draw    (206.98,262.04) -- (206.98,182.44) ;
\draw [shift={(206.98,180.44)}, rotate = 90] [color={rgb, 255:red, 0; green, 0; blue, 0 }  ][line width=0.75]    (10.93,-3.29) .. controls (6.95,-1.4) and (3.31,-0.3) .. (0,0) .. controls (3.31,0.3) and (6.95,1.4) .. (10.93,3.29)   ;
\draw   (10,127.74) -- (69.03,127.74) -- (69.03,186.36) -- (10,186.36) -- cycle ;
\draw (39.52,157.05) node  {\includegraphics[width=44.27pt,height=43.96pt]{ 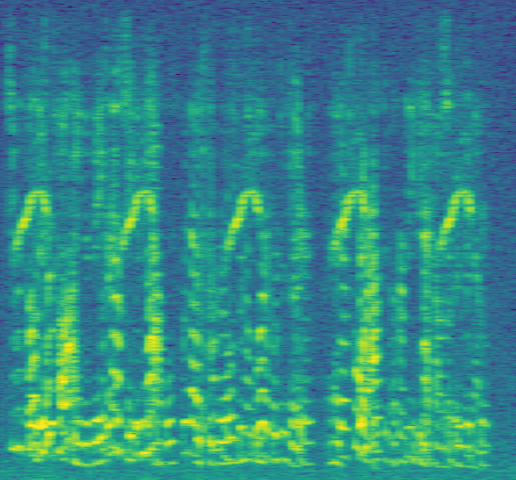}};

\draw   (13,131.14) -- (72.03,131.14) -- (72.03,189.76) -- (13,189.76) -- cycle ;
\draw (42.52,160.45) node  {\includegraphics[width=44.27pt,height=43.96pt]{ images/mix.png}};

\draw   (17,135.14) -- (76.03,135.14) -- (76.03,193.76) -- (17,193.76) -- cycle ;
\draw (46.52,164.45) node  {\includegraphics[width=44.27pt,height=43.96pt]{ images/mix.png}};

\draw   (20,138.54) -- (79.03,138.54) -- (79.03,197.16) -- (20,197.16) -- cycle ;
\draw (49.52,167.85) node  {\includegraphics[width=44.27pt,height=43.96pt]{ images/mix.png}};

\draw   (93.26,134.15) -- (163.36,155.28) -- (163.32,185.82) -- (93.16,206.76) -- cycle ;
\draw    (164.28,170.29) -- (195.19,171.3) ;
\draw [shift={(197.19,171.36)}, rotate = 181.87] [color={rgb, 255:red, 0; green, 0; blue, 0 }  ][line width=0.75]    (10.93,-3.29) .. controls (6.95,-1.4) and (3.31,-0.3) .. (0,0) .. controls (3.31,0.3) and (6.95,1.4) .. (10.93,3.29)   ;
\draw   (197.19,171.36) .. controls (197.19,166.35) and (201.57,162.29) .. (206.98,162.29) .. controls (212.38,162.29) and (216.76,166.35) .. (216.76,171.36) .. controls (216.76,176.37) and (212.38,180.44) .. (206.98,180.44) .. controls (201.57,180.44) and (197.19,176.37) .. (197.19,171.36) -- cycle ; \draw   (200.06,164.95) -- (213.9,177.78) ; \draw   (213.9,164.95) -- (200.06,177.78) ;
\draw   (617.13,206.02) -- (547.02,184.9) -- (547.05,154.36) -- (617.2,133.4) -- cycle ;
\draw   (627.19,138.04) -- (686.23,138.04) -- (686.23,196.66) -- (627.19,196.66) -- cycle ;
\draw (656.71,167.35) node  {\includegraphics[width=44.27pt,height=43.96pt]{ 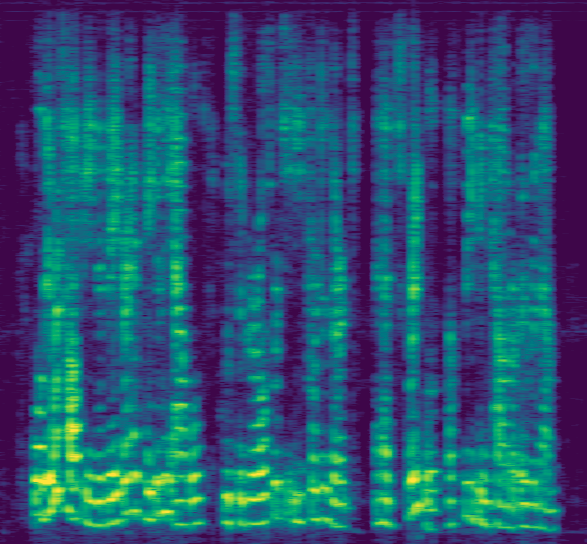}};

\draw    (505.5,170) -- (543.5,170) ;
\draw [shift={(545.5,170)}, rotate = 180] [color={rgb, 255:red, 0; green, 0; blue, 0 }  ][line width=0.75]    (10.93,-3.29) .. controls (6.95,-1.4) and (3.31,-0.3) .. (0,0) .. controls (3.31,0.3) and (6.95,1.4) .. (10.93,3.29)   ;
\draw    (207.5,216.25) -- (441,216.5) -- (441,198.5) ;
\draw [shift={(441,196.5)}, rotate = 90] [color={rgb, 255:red, 0; green, 0; blue, 0 }  ][line width=0.75]    (10.93,-3.29) .. controls (6.95,-1.4) and (3.31,-0.3) .. (0,0) .. controls (3.31,0.3) and (6.95,1.4) .. (10.93,3.29)   ;
\draw    (217.5,170.25) -- (369.5,170.25) ;
\draw [shift={(371.5,170.25)}, rotate = 180] [color={rgb, 255:red, 0; green, 0; blue, 0 }  ][line width=0.75]    (10.93,-3.29) .. controls (6.95,-1.4) and (3.31,-0.3) .. (0,0) .. controls (3.31,0.3) and (6.95,1.4) .. (10.93,3.29)   ;
\draw  [fill={rgb, 255:red, 237; green, 228; blue, 228 }  ,fill opacity=1 ] (400.24,245.35) .. controls (400.24,236.46) and (407.44,229.25) .. (416.34,229.25) -- (591.74,229.25) .. controls (600.63,229.25) and (607.84,236.46) .. (607.84,245.35) -- (607.84,293.65) .. controls (607.84,302.54) and (600.63,309.75) .. (591.74,309.75) -- (416.34,309.75) .. controls (407.44,309.75) and (400.24,302.54) .. (400.24,293.65) -- cycle ;
\draw   (475.74,280.01) -- (504.63,287.53) -- (504.62,297.49) -- (475.7,304.91) -- cycle ;
\draw  [fill={rgb, 255:red, 235; green, 177; blue, 177 }  ,fill opacity=1 ] (506.48,285.76) -- (509.85,285.76) -- (509.85,289.13) -- (506.48,289.13) -- cycle ;
\draw  [fill={rgb, 255:red, 201; green, 135; blue, 135 }  ,fill opacity=1 ] (506.48,289.13) -- (509.85,289.13) -- (509.85,292.5) -- (506.48,292.5) -- cycle ;
\draw  [fill={rgb, 255:red, 184; green, 46; blue, 46 }  ,fill opacity=1 ] (506.48,292.5) -- (509.85,292.5) -- (509.85,295.88) -- (506.48,295.88) -- cycle ;

\draw   (526.46,277.83) .. controls (526.46,275.16) and (528.89,273) .. (531.88,273) .. controls (534.87,273) and (537.3,275.16) .. (537.3,277.83) .. controls (537.3,280.5) and (534.87,282.66) .. (531.88,282.66) .. controls (528.89,282.66) and (526.46,280.5) .. (526.46,277.83) -- cycle ; \draw   (528.05,274.41) -- (535.71,281.24) ; \draw   (535.71,274.41) -- (528.05,281.24) ;
\draw    (470,252) -- (524.46,251.65) ;
\draw [shift={(526.46,251.63)}, rotate = 179.63] [color={rgb, 255:red, 0; green, 0; blue, 0 }  ][line width=0.75]    (10.93,-3.29) .. controls (6.95,-1.4) and (3.31,-0.3) .. (0,0) .. controls (3.31,0.3) and (6.95,1.4) .. (10.93,3.29)   ;
\draw  [fill={rgb, 255:red, 237; green, 189; blue, 189 }  ,fill opacity=1 ] (425.6,242.38) -- (468.41,242.38) -- (468.41,263.69) -- (425.6,263.69) -- cycle ;

\draw    (401,252.8) -- (422.6,252.62) ;
\draw [shift={(424.6,252.6)}, rotate = 179.51] [color={rgb, 255:red, 0; green, 0; blue, 0 }  ][line width=0.75]    (10.93,-3.29) .. controls (6.95,-1.4) and (3.31,-0.3) .. (0,0) .. controls (3.31,0.3) and (6.95,1.4) .. (10.93,3.29)   ;
\draw   (526.46,251.14) .. controls (526.46,248.47) and (528.89,246.31) .. (531.88,246.31) .. controls (534.87,246.31) and (537.3,248.47) .. (537.3,251.14) .. controls (537.3,253.81) and (534.87,255.97) .. (531.88,255.97) .. controls (528.89,255.97) and (526.46,253.81) .. (526.46,251.14) -- cycle ; \draw   (528.05,247.72) -- (535.71,254.55) ; \draw   (535.71,247.72) -- (528.05,254.55) ;
\draw    (532.18,310.4) -- (531.9,284.66) ;
\draw [shift={(531.88,282.66)}, rotate = 89.39] [color={rgb, 255:red, 0; green, 0; blue, 0 }  ][line width=0.75]    (10.93,-3.29) .. controls (6.95,-1.4) and (3.31,-0.3) .. (0,0) .. controls (3.31,0.3) and (6.95,1.4) .. (10.93,3.29)   ;
\draw    (492.37,283.27) -- (492.41,277.83) -- (523.91,277.83) ;
\draw [shift={(525.91,277.83)}, rotate = 180] [color={rgb, 255:red, 0; green, 0; blue, 0 }  ][line width=0.75]    (10.93,-3.29) .. controls (6.95,-1.4) and (3.31,-0.3) .. (0,0) .. controls (3.31,0.3) and (6.95,1.4) .. (10.93,3.29)   ;
\draw    (532,273) -- (531.89,257.97) ;
\draw [shift={(531.88,255.97)}, rotate = 89.6] [color={rgb, 255:red, 0; green, 0; blue, 0 }  ][line width=0.75]    (10.93,-3.29) .. controls (6.95,-1.4) and (3.31,-0.3) .. (0,0) .. controls (3.31,0.3) and (6.95,1.4) .. (10.93,3.29)   ;
\draw    (537.53,251.63) -- (552.14,251.63) ;
\draw [shift={(554.14,251.63)}, rotate = 180] [color={rgb, 255:red, 0; green, 0; blue, 0 }  ][line width=0.75]    (10.93,-3.29) .. controls (6.95,-1.4) and (3.31,-0.3) .. (0,0) .. controls (3.31,0.3) and (6.95,1.4) .. (10.93,3.29)   ;
\draw    (597.51,251.45) -- (618,251.83) ;
\draw [shift={(620,251.87)}, rotate = 181.07] [color={rgb, 255:red, 0; green, 0; blue, 0 }  ][line width=0.75]    (10.93,-3.29) .. controls (6.95,-1.4) and (3.31,-0.3) .. (0,0) .. controls (3.31,0.3) and (6.95,1.4) .. (10.93,3.29)   ;
\draw  [fill={rgb, 255:red, 237; green, 189; blue, 189 }  ,fill opacity=1 ] (555.05,242.29) -- (597.87,242.29) -- (597.87,263.6) -- (555.05,263.6) -- cycle ;

\draw  [dash pattern={on 0.84pt off 2.51pt}]  (377,191.75) -- (406,232.75) ;
\draw  [dash pattern={on 0.84pt off 2.51pt}]  (502,191.75) -- (596.6,229) ;
\draw    (450,264) -- (450,293) -- (472,293) ;
\draw [shift={(474,293)}, rotate = 180] [color={rgb, 255:red, 0; green, 0; blue, 0 }  ][line width=0.75]    (10.93,-3.29) .. controls (6.95,-1.4) and (3.31,-0.3) .. (0,0) .. controls (3.31,0.3) and (6.95,1.4) .. (10.93,3.29)   ;

\draw (104.37,155) node [anchor=north west][inner sep=0.75pt]   [align=left] {{\scriptsize {\fontfamily{pcr}\selectfont Mixture }}\\{\scriptsize {\fontfamily{pcr}\selectfont encoder}}\\};
\draw (465,253) node   [align=left] {\begin{minipage}[lt]{37.65pt}\setlength\topsep{0pt}
{\scriptsize SA}
\end{minipage}};
\draw (57.19,299) node [anchor=north west][inner sep=0.75pt]   [align=left] {{\tiny [T,K,2]}};
\draw (592.21,253) node   [align=left] {\begin{minipage}[lt]{33.78pt}\setlength\topsep{0pt}
{\scriptsize SA}
\end{minipage}};
\draw (53,312) node [anchor=north west][inner sep=0.75pt]   [align=left] {{\fontfamily{ptm}\selectfont {\scriptsize spectral enrollment}}};
\draw (98.19,255) node [anchor=north west][inner sep=0.75pt]   [align=left] {{\scriptsize {\fontfamily{pcr}\selectfont Spectral }}\\{\scriptsize {\fontfamily{pcr}\selectfont encoder}}\\};
\draw (52.7,198) node [anchor=north west][inner sep=0.75pt]   [align=left] {{\tiny [T,K,2J]}};
\draw (389.5,162) node [anchor=north west][inner sep=0.75pt]   [align=left] {{\fontfamily{ptm}\selectfont {\scriptsize Bottelneck proccessing}}};
\draw (663.69,197) node [anchor=north west][inner sep=0.75pt]   [align=left] {{\tiny [T,K,2]}};
\draw (563.69,162) node [anchor=north west][inner sep=0.75pt]   [align=left] {{\scriptsize {\fontfamily{pcr}\selectfont Decoder}}\\\\};
\draw (482,287) node [anchor=north west][inner sep=0.75pt]   [align=left] {{\scriptsize Cls}};
\draw (267,230) node [anchor=north west][inner sep=0.75pt]   [align=left] {{\tiny {\fontfamily{pcr}\selectfont lookup table}}};
\draw (378.81,275.22) node [anchor=north west][inner sep=0.75pt]  [rotate=-179.92]  {$\theta $};
\draw (280,312) node [anchor=north west][inner sep=0.75pt]   [align=left] {{\fontfamily{ptm}\selectfont {\scriptsize spatial enrollment}}};

\end{tikzpicture}
}
\setlength{\belowcaptionskip}{-12pt} 
\caption{Block diagram of the proposed algorithm. The {multiplication} symbol indicates the FiLM operation. SA indicate the self-attention mechanism}
\label{fig:block_diagram}

\end{figure*}

\section{Proposed Model}
\label{sec:model}
In this section, we describe the proposed architecture, the input features, and the training procedure.

\vspace{2pt}\noindent\textbf{Architecture and Features:} In Fig.~\ref{fig:block_diagram}, the proposed algorithm (training and inference stages) is illustrated. The backbone of the proposed method is a U-Net architecture \cite{ronneberger2015u} enhanced with a self-attention mechanism strategically placed at the bottleneck. The mixture encoder comprises \sout{multiple} six convolutional layers, each followed by batch normalization to stabilize training and improve generalization, and a Parametric Rectified Linear Unit (PReLU) activation function \cite{he2015delving} to introduce nonlinearity while avoiding ``dead neuron'' issues. Subsequently, the channel and frequency dimensions are merged, and a fully connected layer is used to reduce dimensionality. A single self-attention layer is then applied. 

The spectral enrollment encoder has the same architecture as the mixture encoder but is adapted to a single-channel input. At its output, the vectors are mean-averaged along the frame dimension to generate a single representation vector, which guides the model toward the target speaker.
The representation of $\td$ is learned using a lookup table. This representation is added to the spectral enrollment embedding to form a unified embedding vector. The summed embedding is then used to condition the mixture embedding by applying  \ac{FiLM} \cite{perez2017film}. This is mathematically defined as:  
\begin{equation}
\scalebox{0.95}{$
    \text{FiLM(x,r)} = \text{emb}_x \cdot \gamma(\text{emb}_r) + \beta(\text{emb}_r),
    $}
\end{equation}
where \(\text{emb}_x\)  represents the mixture embedding, and \(\gamma\) and \(\beta\) are learned feed-forward layers applied to the references embedding \(\text{emb}_r\).
This procedure is applied to each vector along the frame dimension of the mixture embedding on a frame-by-frame basis.

To enable the model to determine which enrollment to prioritize, we introduce a lightweight classifier consisting of three feedforward layers. Operating on the latent representation from the bottleneck layer, this classifier assigns each scenario to one of three categories: (i) both enrollments are relevant, (ii) only the spatial enrollment is relevant, or (iii) only the spectral enrollment is relevant.

{Following the classification, a second pass is performed through the self-attention block. In this phase, the classifier’s output embedding, $\text{emb}_c$, is used to guide the model's attention. Specifically, we apply a \ac{FiLM} operation using both the classifier embedding and the summed enrollment representations, such that the guidance embedding for the second iteration is computed as $\text{FiLM}(r, c)$. This mechanism allows the model to selectively focus on the most relevant enrollment source.
To distinguish between passes, a dedicated prefix token is provided in each iteration, indicating to the self-attention block which iteration is currently being processed.}

The decoder employs transpose-convolution layers to facilitate skip connections between the encoder and decoder. These layers are carefully designed to match the dimensions of the corresponding encoder layers, enabling efficient information transfer and the preservation of fine-grained details during reconstruction. This architecture enables the model to effectively leverage both global context and local details, resulting in enhanced performance.

This work employs the \ac{RI} components of the \ac{STFT} as both input and output features of the model. The real and imaginary parts are merged along the channel dimension. Incorporating \ac{RI} features helps mitigate phase processing challenges.

\vspace{2pt}\noindent\textbf{Training procedure:}
The primary contribution of our work is the integration of two types of enrollment features: spectral and spatial. To ensure robustness, the model must maintain high performance even when one of these features is inaccurate or of low quality. Furthermore, if one enrollment type is accurate but suboptimal for the specific extraction scenario, the model should prioritize the more informative feature. For example, when speakers are located too closely for reliable \ac{DOA}-based separation, the model should favor spectral information to guide the extraction process.
In this context, we define an \textit{erroneous} \ac{DOA} as one that is inaccurately estimated. We model it as a random value between \(0^\circ\) and \(180^\circ\). For spectral enrollment, erroneous data refers to an incorrect speaker reference that does not belong to the desired source. 

To address these challenges, we trained our model using several configurations. Specifically, we sampled a random \ac{DOA}, denoted as $\theta_{\text{rnd}}$, and selected a single-channel reference signal of a random speaker from our corpus who is neither the interfering speaker nor the desired speaker, denoted as $\tilde{e}_{\text{rnd}}$.  The model was then trained using these configurations alongside the nominal configuration, where the \ac{DOA} and spectral enrollments were both correct. It is crucial to ensure that at least one of the enrollments is accurate. The training process is described as follows: 
\begin{subequations}
\begin{align}
   &\scalebox{0.95}{$\hat{\tilde{s}}_d = \mathcal{F}(\mathbf{x}, \tilde{e}_d, \theta_d)$} \label{eq:3a} \\
   &\scalebox{0.95}{$\hat{\tilde{s}}_{d_{\theta_{\text{rnd}}}} =  \mathcal{F}(\mathbf{x}, \tilde{e}_d, \theta_{\text{rnd}})$} &
   \scalebox{0.95}{$\hat{\tilde{s}}_{d_{e_\text{rnd}}} =  \mathcal{F}(\mathbf{x}, \tilde{e}_{\text{rnd}}, \theta_d)$}. \label{eq:3b}
\end{align}
\end{subequations}
We also aim to increase robustness against small errors in the \ac{DOA} estimates. Due to the training process described above, even small \ac{DOA} errors may lead the model to interpret the estimate as a random \ac{DOA}, thereby disregarding its useful spatial guidance. To mitigate this issue, we introduce random perturbations to $\theta_d$, defined as follows: %
\begin{equation}  
\scalebox{0.95}{$
    \theta_d = \theta_d +   \text{Discrete-Uniform}\{\pm 4^\circ, \pm 2^\circ, 0^\circ\}.
    $}
\end{equation}
In each batch, all three scenarios are trained simultaneously, ensuring that the model can adaptively prioritize reliable references without being misled by inaccurate or irrelevant inputs.

We adopted the time-domain \ac{SI-SDR} loss function \cite{le2019sdr} {denoted as $L_{\text{SISDR}}$}, which has proven effective for \ac{BSS} tasks, to train the proposed model to implement the extraction task. 
 %
%
%
%
%
{The classifier was trained using the cross-entropy loss, $L_{\text{CE}}$, which quantifies the difference between its predicted output and the actual setting.}
The loss functions of the three configurations are then summed to obtain the overall loss function:
\begin{equation}
\scalebox{0.95}{$
L_{\text{overall}} = \sum_{a=\{d,d_{\theta_{\text{rnd}}},d_{s_\text{rnd}} \}} 
L_{\text{SISDR}}\left(\tilde{s}_d,\hat{\tilde{s}}_a \right) { + L_{\text{CE}}}.
$}
\end{equation}

\section{Experimental Study}
In this section, we describe the experimental study, including the datasets used, the setup, the inference procedure, and the results. 

\vspace{2pt}\noindent\textbf{Datasets:}
To generate data for training and evaluation, we synthetically created 20,000 training examples, 1,000 validation examples, and 1,000 test examples. The clean speech signals were sampled from the Librispeech dataset, while noise samples were taken from the DNS dataset. The \ac{SNR} was randomly chosen in the range of 5 to 20 dB. For simulating reverberant environments, we used a simulation framework based on the Image method \cite{habets2006room}, where the location and azimuth (with respect to the wall) of the microphone array were randomly assigned for each room. The reverberation time was randomly selected within a range of 0.2 to 0.8 seconds. A \ac{ULA} comprising four microphones, with an 8~cm spacing between adjacent microphones, was used. The minimum distance from the wall was set to 0.5 m. 
Both speakers and the noise were randomly located in the room. The radius of the sources from the array was randomly drawn between 1 and 4 m. The \ac{DOA} of the sources was measured relative to the center of the array and was selected with a resolution of $2^\circ$ in the range of \(0^\circ\) to \(180^\circ\). 

\vspace{2pt}\noindent\textbf{Test Configurations:}
To thoroughly evaluate our model under various challenging scenarios, we created six types of  test configurations:  
1)~Close Spatial Proximity (CSP): The speakers in the mixture are separated by only 2 degrees, with differing radii to ensure they are not located in the same spot;  
2)~Moderate Spatial Proximity (MSP): The speakers are separated by $10^\circ$, a close yet sufficient distance for a spatial-based model to effectively extract the desired speaker; 3)~Same-Gender Mixtures (SGM): The speakers in each mixture have the same gender and are randomly positioned, making it more challenging for spectral-based models to distinguish between them; 4)~Random \ac{DOA} Reference (SGM-RDR): Similar to Scenario 3, but with random \ac{DOA} values provided as references to test the model's robustness to incorrect directional information; 5)~Random Spectral Reference (SGM-RSR): Similar to Scenario 3, but with a randomly selected spectral reference that does not correspond to any speaker in the mixture, ensuring the reference is unrelated to the mixture's participants; and 6)~Low-SNR Spectral Enrollment (SGM-LSSE): Similar to Scenario 3, but the spectral enrollment is corrupted by additive white noise with an SNR value randomly sampled between -2 dB and 2 dB. 

\vspace{2pt}\noindent\textbf{Algorithm Settings:}
The speech and noise signals were first downsampled to 8~kHz. The \ac{STFT} is calculated with a frame size of 256 samples, $50\%$ overlap, and a Hamming window. Due to the symmetry of the \ac{DFT}, only the first 129 frequency bins are processed.  
For training, the AdamW optimizer~\cite{loshchilov2017decoupled} is used with a learning rate of 0.0001 and a batch size of 14. The model weights are initialized randomly, and the length of the signals is set to 4~Sec.

\begin{table*}[ht]
\begin{center}
\setlength{\belowcaptionskip}{-32pt}
\caption{SI-SDRi results in dB of the proposed model's variants for various test sets. The label \( \times \) indicates that the results are irrelevant. ``-w'' and ``-w/o'' stand for ``with'' and ``without,'' respectively.}
\label{table:results}
\renewcommand{\arraystretch}{0.85} 
\small 

\scalebox{0.98}{ 
\begin{tabular}{@{}lccccccc@{}}
\toprule
\textbf{Test Set / Model} & \textbf{Unprocessed} & \textbf{Spectral-only} & \textbf{Spatial-only} & \textbf{Proposed} & \textbf{Proposed w/o \eqref{eq:3b}} & \textbf{Proposed w. \ac{DOA} inference} \\ 
\midrule
\addlinespace
\textbf{CSP} & -0.76 & 6.85 & -3.36 & \textbf{7.58} & 5.9 & 6.4 \\
\textbf{MSP} & -0.79 & 8.14 & 7.19 & \textbf{10.3} & 9.73 & 7.51 \\
\textbf{SGM} & -0.77 & 6.83 & 8.33 & 9.58 & \textbf{9.61} & 7.95 \\
\textbf{SGM-RDR} & -0.77 & 6.83 & \( \times \) & \textbf{7.8} & -3.41 & \( \times \) \\
\textbf{SGM-RSR} & -0.77 & \( \times \) & 8.33 & \textbf{8.86} & 7.01 & 5.23 \\
\textbf{SGM-LSSE} & -0.77 & -2.08 & 8.33 & \textbf{9.24} & 7.6 & 5.48 \\
\bottomrule
\end{tabular}
} 

\end{center}
\vspace{-.8cm}
\end{table*}

\vspace{2pt}\noindent\textbf{Inference Procedure:}
Enrollment is required to guide the model toward the desired speaker, either through a single-channel reference signal or a \ac{DOA}. While the single-channel reference is assumed to be available a priori, the \ac{DOA} can be estimated directly from the mixture. Previous works often assume that the \ac{DOA} is known in advance or obtained from an external modality, such as a camera. As this assumption can be restrictive, we propose a method to estimate the \ac{DOA} independently.
To this end, we trained a compact network mirroring the mixture encoder architecture, augmented with a fully connected layer that projects the embedding into a 91-dimensional \ac{DOA} space. We optimized this network with a binary cross-entropy loss to predict only the two speech sources’ \acp{DOA} present in the mixture, explicitly learning to ignore the directional noise. 

Once both \acp{DOA} are estimated, the next step is to determine which \ac{DOA} corresponds to each spectral enrollment.
Assuming that our speaker extraction model is robust enough to accommodate erroneous enrollments, it can facilitate the following matching process. 
Using two spectral and two spatial enrollments (one for each speaker), we extract four signals, each inferred with one correct enrollment and one intentionally erroneous. Specifically, for spectral-only enrollment, we substitute the true \ac{DOA} with a random \ac{DOA} to obtain \(\hat{\tilde{s}}_{\text{ref}_q}\) for \(q=1,2\), and for spatial-only enrollment, we replace the speech reference with Gaussian noise to extract \(\hat{\tilde{s}}_{\theta_q}\) for \(q=1,2\).
Finally, using the \ac{SI-SDR} metric, we determined the optimal pairing between the estimated \acp{DOA} and the spectral enrollments, ensuring accurate speaker assignment (recall that $Q=2$):
\begin{equation}
\scalebox{0.95}{$
\sum_{q=1}^{2} \text{SI-SDR}(\hat{\tilde{s}}_{\theta_q}, \hat{\tilde{s}}_{\text{ref}_q})  
\lessgtr 
\sum_{q=1}^{2} \text{SI-SDR}(\hat{\tilde{s}}_{\theta_q}, \hat{\tilde{s}}_{\text{ref}_{3-q}})
$}.
\end{equation}

\vspace{2pt}\noindent\textbf{Results:}
The results of the \ac{SI-SDR} improvement are depicted in Table~\ref{table:results} for all six test sets.  
We evaluate five variants of our model:  
1) Spectral-only, trained solely on spectral information, with the enrollment being a single microphone signal of the desired speaker; 
2) Spatial-only, trained exclusively on spatial information based on DOA;  
3) The proposed model; 
4) Our proposed model, assuming the \ac{DOA} is always accurate, i.e.~it is trained without using \eqref{eq:3b}; and 
5) The proposed model with \ac{DOA} estimation.  
The first two variants highlight the main challenges. The spectral-based model displays relatively stable performance even when the speakers are in close proximity. However, some degradation is observed in these conditions because spatial cues in the mixture can still affect the output. Conversely, the DOA-based model completely collapses when the speakers are close but performs better when they share the same gender, provided their \ac{DOA} are further apart. It is also crucial to note that if either model encounters errors in the enrollment signal or the enrollment suffers from poor quality, as seen in the SGM-LSSE configuration, it fails to perform speaker extraction due to the absence of a reliable alternative reference.

By using both enrollments without the training procedure described in \eqref{eq:3b}, the model primarily relies on spatial information, suggesting that spatial cues are more “intuitive” for separation tasks. {Consequently, performance remains relatively stable when the spectral reference is inaccurate, with only a 2~dB degradation compared to the case of an accurate spectral reference. In contrast, when the \ac{DOA} is estimated incorrectly, the model’s performance deteriorates much more severely, ultimately resulting in a failure of separation.}

In the proposed model (with a classifier), both references are leveraged effectively. Even if one of the references, either spectral or spatial, is inaccurate, the model remains stable in extracting the target speaker. However, when the two mixed speakers are very close, the spatial information can introduce additional complexity, leading to some performance degradation. Despite this, the overall separation capabilities remain relatively robust.
Finally, we evaluated the model using our \ac{DOA} inference procedure. Since these \ac{DOA} estimates can be inaccurate, performance degradation is less severe when the model relies primarily on spectral information rather than spatial cues, as is often the case when speakers are close together. Conversely, in scenarios where spatial cues are more informative (i.e., when speakers are well separated), errors in the estimated \acp{DOA} lead to a more pronounced drop in performance.

\begin{figure}[htbp]
    \centering
    \vspace{-.25cm}
    \includegraphics[width=1.0\linewidth,height=6.5cm]{ 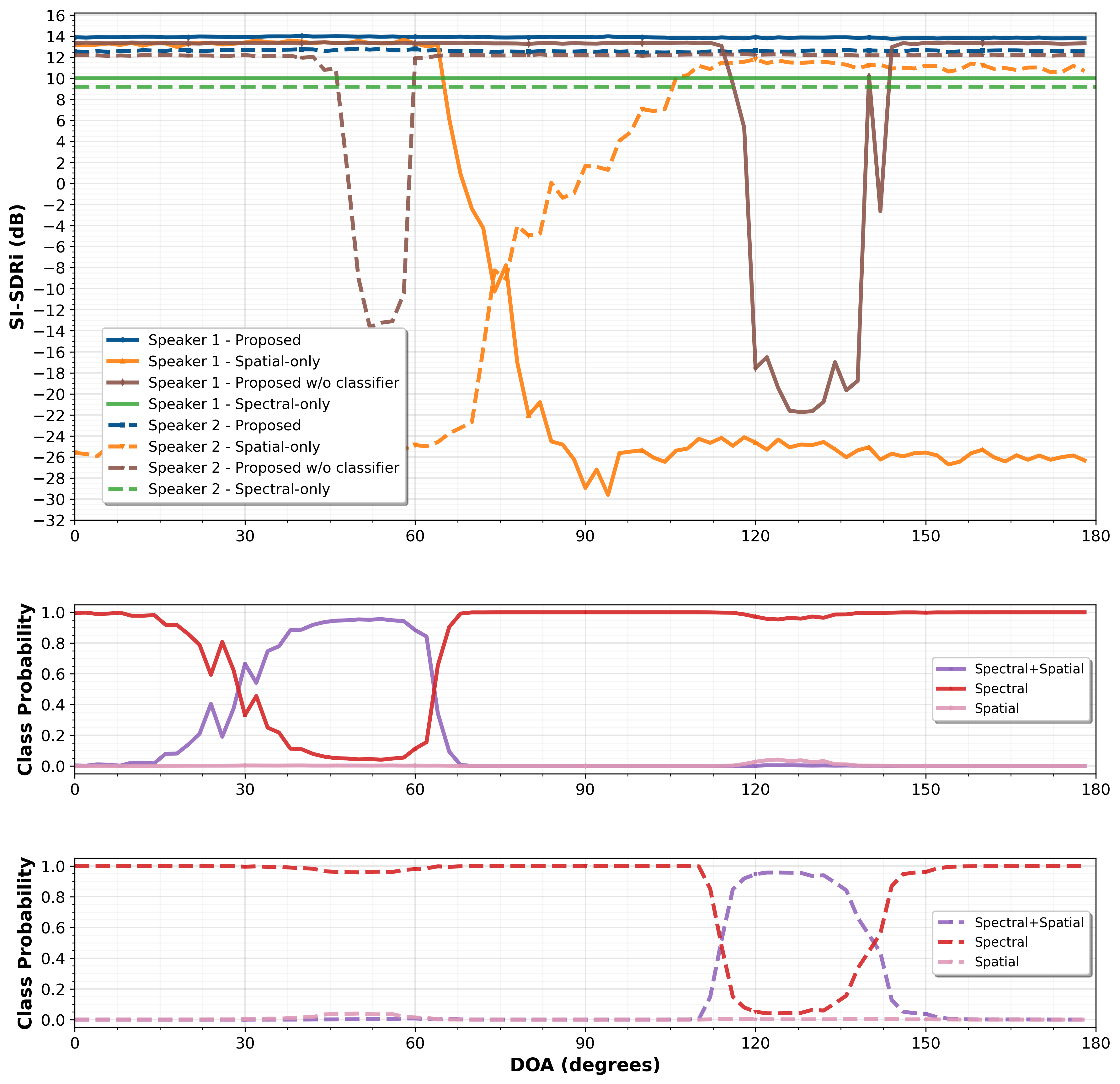}
    \addtolength{\belowcaptionskip}{-25pt}
    \addtolength{\abovecaptionskip}{-12pt}\caption{Performance and classifier analysis of the proposed model as a function of {the enrollment} angle $\theta$ (in degrees) where the first speaker was positioned at 54° and the other at 122°. \textbf{Top}: SI-SDRi comparison for {four} configurations — Spatial-only, {Spectral-only,} and Proposed (w.~and w/o classifier) — {evaluated on the two-speaker case.} \textbf{Middle and Bottom}: Classifier output probabilities from the Proposed model indicating the decision confidence as a function of $\theta$ for the first and second speakers.}\label{fig:sisdr_vs_theta}
\end{figure}

To evaluate the robustness of the proposed model against inaccurate spatial enrollment, we compared it with the spectral-only and spatial-only baselines, as well as a variant of the proposed model without the classifier. In this experiment, one speaker was placed at $54^\circ$ and the other at $122^\circ$ relative to the array center, under a fixed \ac{SNR} of 20~dB. The methods were then tested using \ac{DOA} enrollments spanning the range $0^\circ$–$180^\circ$, i.e., the spatial information could be erroneous, while the correct spectral enrollment was consistently provided. 
The results are depicted in Fig.~\ref{fig:sisdr_vs_theta}. It can be clearly observed that the spectral-only model, which does not depend on the \ac{DOA}, maintains a stable \ac{SI-SDR} for both speakers across all angular configurations. In contrast, the spatial-only model performs well only when the given \ac{DOA} closely matches the actual target speaker and is sufficiently separated from the interferer, but its performance degrades sharply when this condition is not satisfied, especially when the spatial enrollment is associated with the wrong speaker. 
The full proposed method achieves robust source extraction even when the spatial enrollment corresponds to the interfering speaker. However, if the classifier is excluded from the model, spatial enrollments associated with the wrong speaker lead to degraded performance. This highlights the ability of the full model (with the classifier) to down-weight misleading spatial cues and rely more heavily on spectral information under adverse conditions, thereby preserving extraction quality.
To further interpret the role of the classifier, the class probabilities for the three categories are shown in the middle and bottom panels of Fig.~\ref{fig:sisdr_vs_theta}. The results demonstrate that the classifier successfully learned to predict the correct class for each enrollment condition.

\section{Conclusions}
{In this paper, we presented a multi-channel speaker extraction model that integrates spatial and spectral enrollments through a scenario-classification learning module and a joint training strategy. The proposed approach effectively balances these two types of references, ensuring stable performance even when one is inaccurate. Our experiments confirm that the model consistently extracts the target speaker under challenging conditions, including degraded \ac{DOA} enrollments or unreliable single-channel spectral references. These findings highlight the robustness and practical applicability of the method for real-world speaker extraction scenarios.

\balance
\bibliographystyle{IEEEtran}

\bibliography{ main}

\end{document}